\documentclass[a4paper]{jpconf}
\usepackage{graphicx}
\begin{document}
\title{Weak Localisation in Clean and Highly Disordered Graphene}

\author{Michael Hilke, Mathieu Massicotte, Eric Whiteway, Victor Yu}

\address{Department of Physics, McGill University, Montr\'eal, Canada H3A 2T8}
\ead{hilke@physics.mcgill.ca}

\begin{abstract}
We look at the magnetic field induced weak localisation peak of graphene samples with different mobilities. At very low temperatures, low mobility samples exhibit a very broad peak as a function of the magnetic field, in contrast to higher mobility samples, where the weak localisation peak is very sharp.  We analyze the experimental data in the context of the localisation length, which allows us to extract, both the localisation length and the phase coherence length of the samples, regardless of their mobilities. This analysis is made possible by the observation that the localisation length undergoes a generic weak localisation dependence with striking universal properties.
\end{abstract}

\section{Introduction:} Weak localisation (WL), which is the enhanced backscattering of coherent electrons in a disordered media \cite{Hikami80,Altshuler80}, has been observed in many materials, including in thin metals and in two dimensional electron gases \cite{Bergman84}. Since WL strongly depends on phase coherence, it is often used to probe the phase coherence properties of a device and as a tool for extracting information on various scattering mechanisms. More recently, WL was discussed and observed in graphene \cite{Ando02,McCann06,Morozov08}, where the existence of two degenerate Dirac valleys leads to interesting new WL properties. For instance, the strength of the WL effect strongly depends on inter-valley scattering, and WL is expected to vanish in its absence \cite{Cooper12}. Perturbative expressions for the conductivity corrections in graphene due to WL have been worked out in great detail by McCann et al. \cite{McCann06} and successfully fitted to the magnetic field dependence of the resistance at low temperatures \cite{Morozov08}. The dominant feature is a peak in the resistance at zero magnetic field. Most of the works were done in a regime of weak disorder, where the perturbative approach by McCann et al. is applicable and where strong localisation is absent.

The aim of this paper is to discuss an alternate point of view to WL, which is not restricted to weak disorder. This was largely driven by the desire to understand experimental observations of WL in a large range of graphene samples, of which some were highly disordered. At low temperatures, the sharpness of the WL peak in clean large scale graphene is quite striking \cite{Whiteway10}, whereas in low mobility graphene samples, a wide negative magnetoresistance behaviour is observed \cite{zhu11}. What we show in this work, is that WL can be naturally understood within the framework of strong localisation and applied to the experimental case of strong disorder as well as weak disorder.

\section{Localisation length:} Our results follow from the observation shown in figure 1, where the localisation length ($L_c$) exhibits a dip at zero magnetic field, which in turn leads to a peak in the resistance. $L_c$ was extracted by computing the resistance of a disordered graphene flake using non-equilibrium Green's functions. The two terminal resistance is obtained by computing the Green's function of the honeycomb tight binding lattice iteratively as a function of the length of the system. $L_c$ is then determined from the exponential increase in resistance with the length of the graphene flake, when the length exceeds $L_c$. Disorder is assumed to be on-site and uncorrelated. Figure 1 shows the dip at zero field of $L_c$ for different strengths of the disorder $V$, where each carbon site has a random potential $-V/2<v_i<V/2$. $V$ is in units of the hopping element, which is $\sim 3eV$ in graphene.

\begin{figure}[!h]
\begin{minipage}{18pc}

\includegraphics[width=18pc]{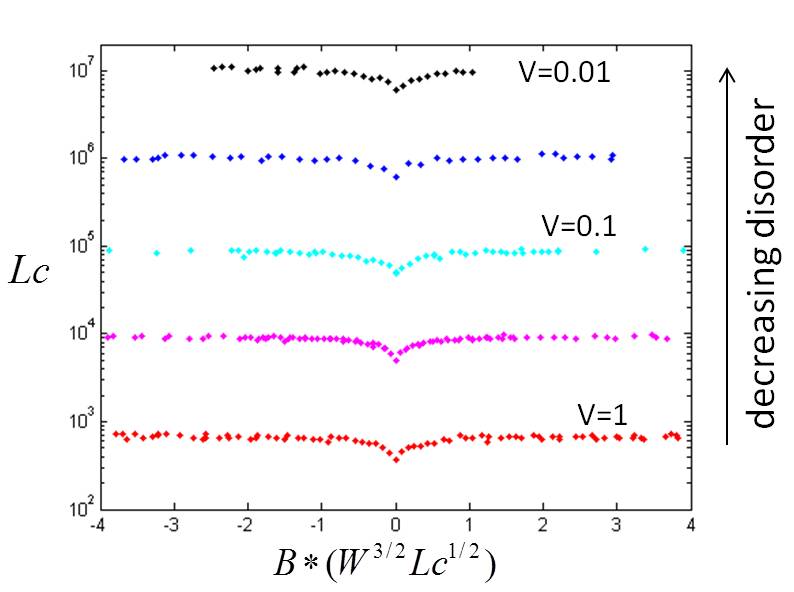}
\caption{The magnetic field dependence of the extracted localisation length (in units of lattice sites) for different values of the disorder $V$.}
\end{minipage}\hspace{2pc}
\begin{minipage}{18pc}
\includegraphics[width=18pc]{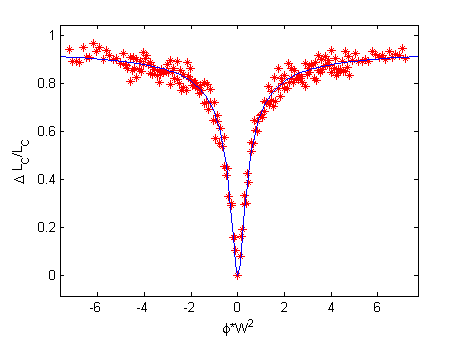}
\caption{The magnetic field dependence of the relative localisation length for the disorder strength $V=0.3$ and for different ribbon widths (40 to 120 lattice sites). The line is fitted according to equation (1).}
\end{minipage}
\label{numeric}
\end{figure}

$L_c$ follows the same behavior regardless of the amount of disorder or the width of the device. This is nicely illustrated by rescaling the magnetic field by $B\sqrt{W^3Lc}$ and then by plotting the relative $\Delta L_c/L_c=(L_c(B)-L_c(0))/L_c(0)$ as a function of field. In this case all curves overlap and are described by the same function within numerical accuracy. The edge plays an important role here because, in the  weak disorder limit, we always have $L_c\gg W$, where $W$ is the width of the system. This is because $L_c$ scales as $L_c\sim W$ for a given disorder strength, hence no matter how wide the device, we will always have $L_c\gg W$ in the low disorder limit.

In the presence of edge scattering, it was shown that the correction to the conductivity is given by $\Delta\sigma\sim (1-1/\sqrt{1+B^2/B_c^2})$, where $B_c$ is a characteristic field determined by the device width and the scattering strength \cite{McCann06}. Since Anderson localisation is also a measure of the return probability, it is not surprising that the same functional form also determines $L_c$, hence we find that the expression

\begin{equation}
\Delta L_c/L_c=\frac{L_c(B)-L_c(0)}{L_c(0)}\simeq 0.95\left(1-1/\sqrt{1+B^2/B_c^2}\right)
\end{equation}

convincingly fits the numerical data shown in figure 2. However, when the disorder becomes very strong as measured by $L_c\ll W$, edge scattering is no longer relevant, since in this case $W$ always exceeds $L_c$. This becomes then a two-dimensional problem, where the shortest length scale is given by $L_c$. In this case expression (1) is not valid anymore and we have to use a two-dimensional analogue. Numerically, this is illustrated in figure 3, where we show again the relative $L_c$, as a function of magnetic field, but this time for stronger disorder. When comparing figure 3 to figure 2, the shape has changed but not the general behavior, namely a dip at zero field.

\begin{figure}[!h]
\begin{minipage}{18pc}

\includegraphics[width=18pc]{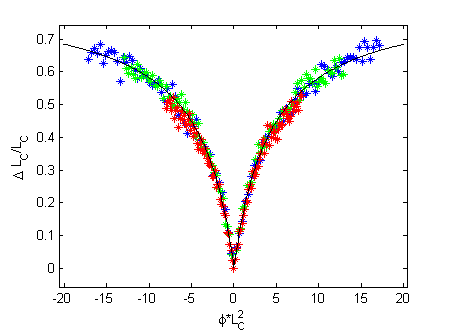}
\caption{The magnetic field dependence of the relative $L_c$ for the disorder strength $V=2$ and for different widths (40 to 120). The full line is obtained using equation (2).}
\end{minipage}\hspace{2pc}
\begin{minipage}{18pc}
\includegraphics[width=18pc]{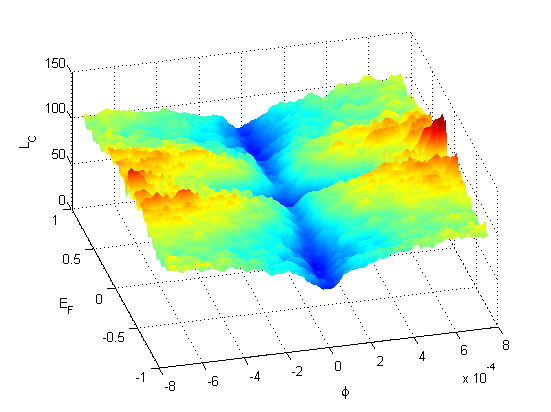}
\caption{The density (or Fermi energy) dependence, as well as the magnetic field dependence of $L_c$ is plotted for $V=2$ and the width $W=120$. The Dirac point corresponds to $E_F=0$.}
\end{minipage}
\label{numeric}
\end{figure}

The functional form of the field dependence has to be modified according to the two-dimensional nature of the problem and can be obtained by following, for instance, Beenakker's approach to WL \cite{Beenakker88}. The field correction to the conductivity can be written as $\Delta\sigma\sim\ln(t_\phi/\tau_e+1)$, where $t_\phi\sim 1/(1/\tau_\phi+1/\tau_B)$, $\tau_e$ is the elastic scattering time, $\tau_\phi$ the phase coherence time and $\tau_B\sim B^{-1}$ the field induced destruction of coherent backscattering. For the numerical simulations, we effectively work at zero temperature, which leads to an infinite phase coherence time. In this case, the coherent backscattering time is not bound by $\tau_\phi$ but is bound by the localisation time, $\tau_c$, where $L_c=\sqrt{D\tau_c}$ and $D$ is the diffusion constant. Hence, for the simulations, $\tau_\phi$ has to be replaced by $\tau_c$ in the expression for the field dependence. Now, assuming again that the relative change in $L_c$ is proportional to the coherent return probability, i.e., the relative change of the conductivity and further assuming that $\tau_e\simeq\tau_c$, we obtain


\begin{equation}
\Delta L_c/L_c\simeq 1.4\cdot\ln\left(\frac{1+B/B_c}{1+B/2B_c}\right),
\end{equation}

\noindent which nicely fits the numerical data in figure 3 (the prefactor is $\sim 1/\ln(2)$ which normalizes expression (2) at high fields). We used $\tau_c/\tau_B=B/B_c$, where  $B_c$ is again the characteristic magnetic field. In the limit, where $\tau_e\ll\tau_c$, we obtain simply $\Delta L_c/L_c\sim\ln(1+B/B_c)$, which leads to a renormalization of $B_c$ by a factor of 2 at small fields. At low temperatures and in the presence of strong disorder, $L_c$ is the shortest length scale and the characteristic field is $B_c\simeq\hbar/eL_c^2$, which defines the area of a phase coherent loop, before coherent backscattering is suppressed when penetrated by a flux quantum. Indeed, equation (2) fits the numerical results for the relative $L_c$, when the field is rescaled by $B\cdot L_c^2$, as shown in figure 3. For small disorder, on the other hand, where $L_c\gg L_\phi$, the characteristic field is instead determined by $L_\phi$ and given by $B_c\simeq\hbar/eL_\phi^2$. Hence, using equation (2) we can describe the generic behaviour of both the low disorder ($L_c\gg L_\phi$) and high disorder ($L_c\ll L_\phi$) limits. To make the connection to localisation as above, it is important to distinguish between the behaviour of $L_c$ and the behaviour of the relative $L_c$. Indeed, $L_c$ does not follow a generic dependence on field or density as shown in figure 4. It is only the relative $L_c$ which shows this generic and an almost universal behaviour as graphed in figure 8.

\section{Experiments:}
The next step is to apply these concepts to the experimental data. We performed experiments on large scale graphene as well as lithographically defined Hall bars and graphene nano-ribbons, in addition to large (over 100 $\mu$m) single crystal grains, of which some show strong dendritic structures and are dubbed graphlocons. Monolayers of graphene were grown by chemical vapor deposition (CVD) of hydrocarbons on 25 $\mu$m-thick commercial Cu foils. The CVD process used was similar to those  described in previous works \cite{Whiteway10,Yu11,Bernard11}.

\begin{figure}[!h]
\begin{minipage}{18pc}

\includegraphics[width=18pc]{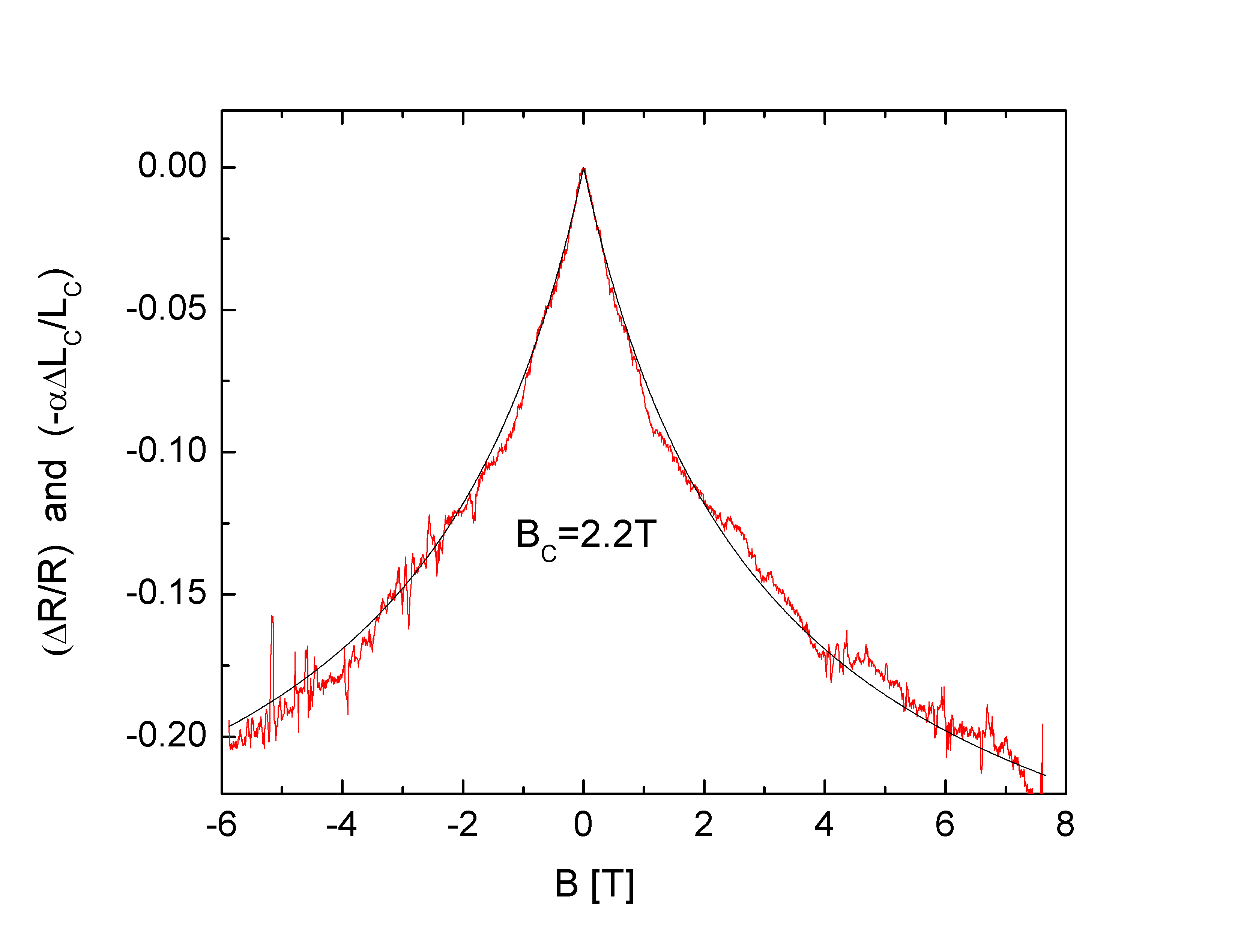}
\caption{The measured magneto-resistance of a low mobility ($\mu\simeq 100 cm^2/Vs$) graphene Hall bar device at 100mK. Shown is the relative resistance and the fit to the relative $L_c$. The fitting parameters are $B_c=2.2$T and $\alpha=0.4$.}
\end{minipage}\hspace{2pc}
\begin{minipage}{18pc}
\includegraphics[width=18pc]{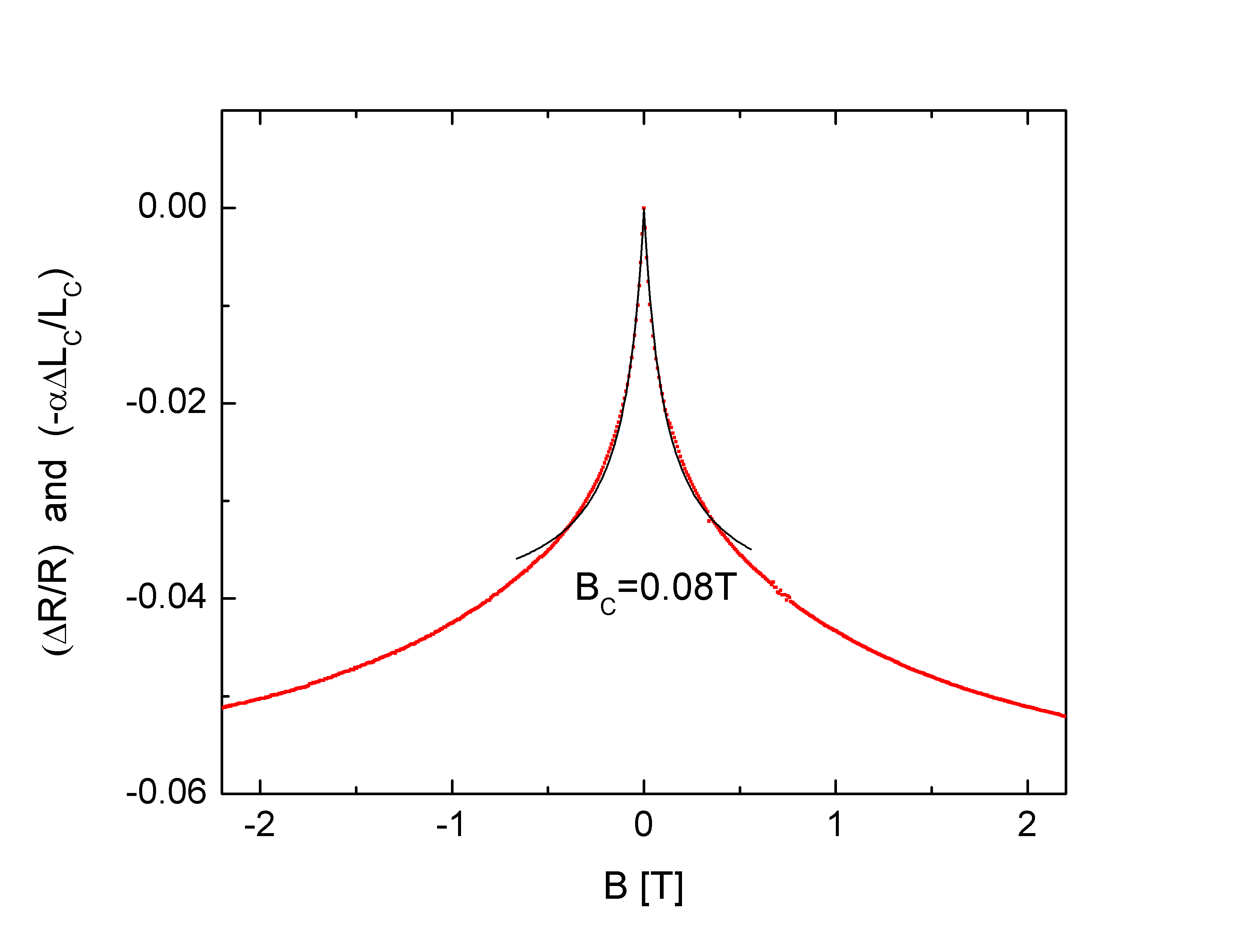}
\caption{This sample had a medium mobility of $\mu\simeq 700 cm^2/Vs)$ and the fitting parameters are $B_c=0.08$T and $\alpha=0.06$.}
\end{minipage}
\label{numeric}
\end{figure}

The overall experimental behavior can be summarized as follows: at low temperatures, low mobility samples, typically below 500 cm$^2$/V$\cdot$s show a very wide peak in the resistance at zero field (as shown in figure 5). With increasing mobility, the peak becomes increasingly sharper (figures 6 and 7). All the samples we measured show a WL peak at zero field. The relative change in resistance varies between 0.5\% for high mobilities to 20\% for low mobilities. This is also expected from McCann's expression, since it yields $\Delta \rho/\rho\sim\rho\Delta F(B)$, where $\Delta F(B)$ describes the magnetic field dependence and $\rho$ is the resistivity \cite{McCann06}. The main difference is that equation (2) leads to an increased sensitivity on the sample geometry, since for a phase coherent sample the resistance will depend exponentially on the length at lengths longer than the localization length.

To connect the data with $L_c$, we can simply assume that the resistance is given by $R\sim e^{L/L_c}$ where $L$ is the length of the sample. If $L_\phi$ is smaller than the length of the sample, then $L$ needs to be replaced by $L_\phi$. For small changes in the relative resistance, we can then write

\begin{equation}
\frac{\Delta R}{R}\simeq-\frac{\Delta L_c}{L_c}\cdot \frac{L_\phi}{L_c}.
\end{equation}

Using equations (2) and (3) we can now fit the relative resistance as a function of the magnetic field for different samples. The fits are shown in figures 5 to 7 for samples with increasing mobilities. The fits are quite good and require only two fitting parameters, $B_c$ and $\alpha\simeq 1.4 L_\phi/L_c$, which are given in the corresponding figure captions. Recalling that for large devices, the limits are given by $B_c\simeq\hbar/e\min\{L_\phi^2,L_c^2\}$, which can be interpolated to $B_c\simeq\hbar(L_\phi^{-2}+L_c^{-2})/e$ and which allows us to obtain $L_c$ and $L_\phi$. Plugging in the numbers we find that $L_\phi\simeq 20$nm and $L_c\simeq 60$nm for the sample in figure 5, whereas $L_\phi\simeq 90$nm and $L_c\simeq 2\mu$m for the sample in figure 6, and finally $L_\phi\simeq 0.5\mu$m and $L_c\simeq 65\mu$m for the sample in figure 7.

\begin{figure}[!h]
\begin{minipage}{18pc}

\includegraphics[width=18pc]{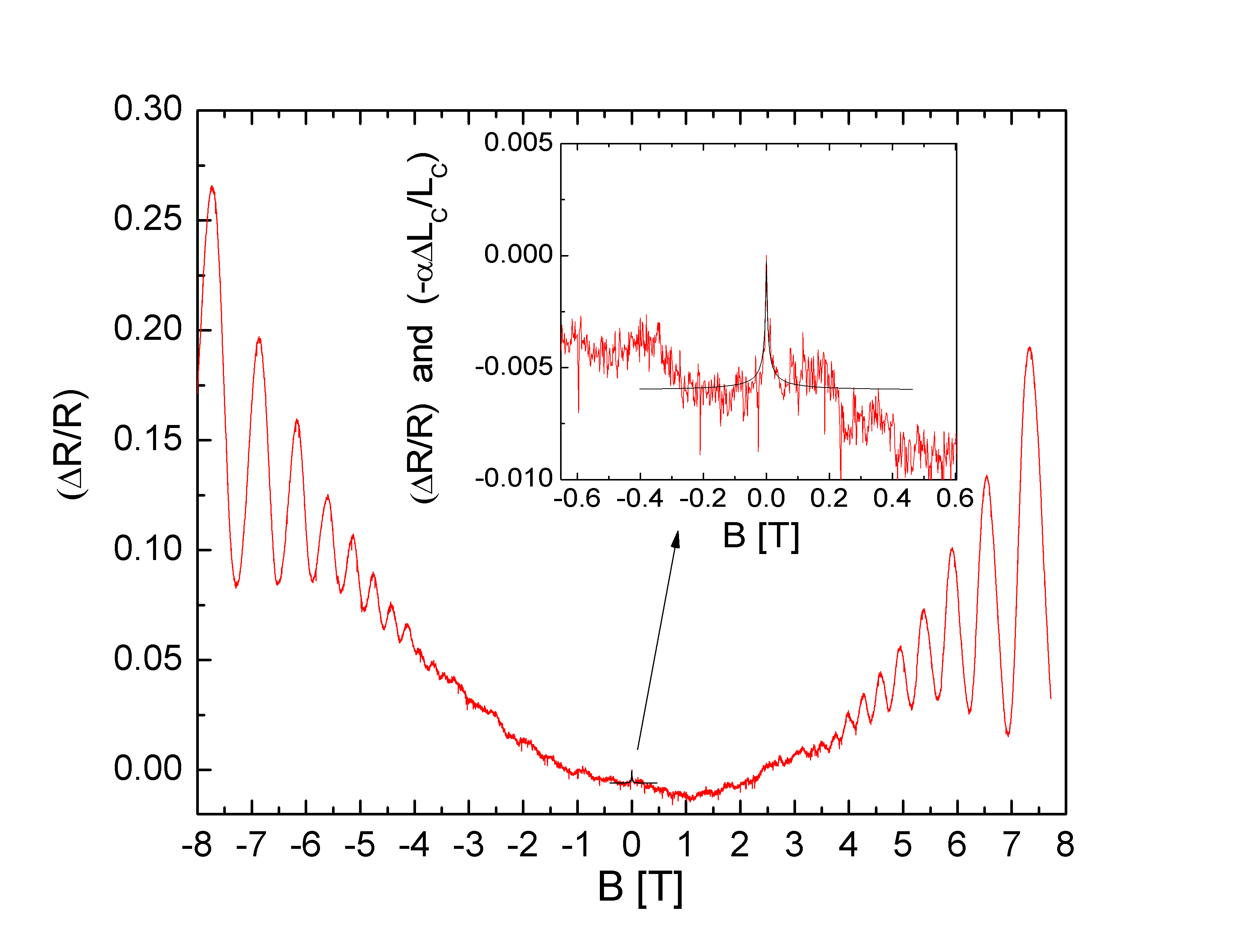}
\caption{The relative resistance change of a high mobility sample ($\mu\simeq 5000 cm^2/Vs)$ with fitting parameters $B_c=0.003$T and $\alpha=0.01$. The inset is a zoom-in of the low field region.}
\end{minipage}\hspace{2pc}
\begin{minipage}{18pc}
\includegraphics[width=18pc]{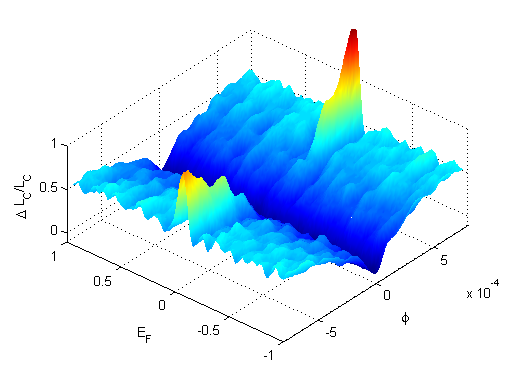}
\caption{The relative $L_c$ as a function of the Fermi energy and magnetic field for $V=2$ and $W=120$. The absence of a systematic dependence on the Fermi energy outside of the Dirac peak is quite striking and determines a close to universal dependence, irrespective of density and disorder strength.}
\end{minipage}
\label{numeric}
\end{figure}

\section{Conclusion:}
We presented an alternate way of looking at the experimental weak localisation peak, by realizing that the localisation length also undergoes a weak localisation effect, in the sense that there is a dip at zero field, which implies a peak in resistance. This allows us to connect weak localisation to strong localisation in a natural way. Moreover, we can associate a characteristic field, $B_c$, which determines the magnetic field dependence. $B_c$ is determined by the square of the inverse localization length for strong disorder and by the square of the inverse phase coherence length for weak disorder. Indeed, we find that the magnetic field dependence of the relative localisation length becomes close to universal, and that all curves can be rescaled on top of each other, by rescaling the field dependence with $B_c$. We used this generic localisation length dependence to fit our experimental data for graphene samples of very different mobilities, which not only allowed us to obtain a good fit, but also to extract the localisation length and phase coherence length independently. This method is very robust and is applicable to high and low mobility samples. We expect this analysis to be very generic and applicable to a large class of two dimensional systems and not only to graphene.

\section{Acknowledgments:}
We thank R. Gagnon and J. Lefebvre for technical assistance and NSERC and FQRNT for financial assistance.

\end{document}